\documentclass[aps,prd,twocolumn,
amssymb,amsmath,showpacs,a4paper,superscriptaddress,nofootinbib,longbibliography]{revtex4-2}
\usepackage{graphicx}
\usepackage{braket}
\usepackage{amsfonts}
\usepackage{amsmath}
\usepackage{bbold}
\usepackage{hyperref}
\usepackage{adjustbox}
\usepackage{float}
\usepackage[nameinlink]{cleveref}
\usepackage[table,xcdraw]{xcolor}
\usepackage{mathrsfs} 

\newcommand{\beq}{\begin{equation}}
	\newcommand{\eeq}{\end{equation}}
\newcommand{\bea}{\begin{eqnarray}}
	\newcommand{\eea}{\end{eqnarray}}
\newcommand{\bit}{\begin{itemize}}
	\newcommand{\eit}{\end{itemize}}
\newcommand{\ben}{\begin{enumerate}}
	\newcommand{\een}{\end{enumerate}}
\newcommand{\nn}{\nonumber}

\def\scri{\mathscr{I}}
\def\H{\mathscr{H}}

\begin{document}

\title{Pseudospectrum of rotating analog black holes}

\author{Lucas Tobias de Paula}
	\email{tobias.l@ufabc.edu.br}
        \affiliation{Centro de Matem\'atica, Computa\c c\~ao e Cogni\c c\~ao, Universidade Federal do ABC (UFABC), 09210-170 Santo Andr\'e, S\~ao Paulo, Brazil}
	\affiliation{Centro de Ci\^encias Naturais e Humanas, Universidade Federal do ABC (UFABC), 09210-170 Santo Andr\'e, S\~ao Paulo, Brazil}
\author{Pedro Henrique Croti Siqueira}
	\email{pedro.croti@ufabc.edu.br}
       \affiliation{Centro de Matem\'atica, Computa\c c\~ao e Cogni\c c\~ao, Universidade Federal do ABC (UFABC), 09210-170 Santo Andr\'e, S\~ao Paulo, Brazil}
	\affiliation{Centro de Ci\^encias Naturais e Humanas, Universidade Federal do ABC (UFABC), 09210-170 Santo Andr\'e, S\~ao Paulo, Brazil}
	\author{Rodrigo Panosso Macedo}
	\email{rodrigo.macedo@nbi.ku.dk}
	\affiliation{Center of Gravity, Niels Bohr Institute, Blegdamsvej 17, 2100 Copenhagen, Denmark}
	\author{Maur\'icio Richartz}
	\email{mauricio.richartz@ufabc.edu.br}
	\affiliation{Centro de Matem\'atica, Computa\c c\~ao e Cogni\c c\~ao, Universidade Federal do ABC (UFABC), 09210-170 Santo Andr\'e, S\~ao Paulo, Brazil}

\begin{abstract}
   Analyzing the stability of quasinormal modes (QNM) is essential for understanding black hole dynamics, particularly in the context of gravitational wave emissions and black hole spectroscopy. In this study, we employ the hyperboloidal approach to reformulate the quasinormal mode problem of a rotating analog black hole, effectively transforming it into an eigenvalue problem associated with a nonself-adjoint operator. Using this method, we examine the influence of rotation on the stability of the QNM spectrum, relying on the associated pseudospectrum for qualitative assessment. 
   Our findings indicate that the prograde overtones become more stable as rotation increases. 
   This work enhances our understanding of spectrum stability in rotating systems and expands the study of pseudospectra in non-spherically symmetric spacetimes, with potential for empirical testing in terrestrial laboratories.
  
\end{abstract}

\maketitle
\let\originalnewpage\newpage \let\newpage\relax \let\newpage\originalnewpage

\section{Introduction}
\label{introd}

Gravitational waves (GWs), predicted by Einstein’s theory of general relativity and recently detected in terrestrial laboratories by the LIGO, Virgo, and KAGRA collaborations~\cite{ligo16,KAGRA:2021vkt,LIGOScientific:2024elc}, provide direct observations of some of the universe’s most extreme phenomena, thereby enhancing our understanding of gravitational systems. Among the signatures embedded within GW signals are quasinormal modes (QNMs), which arise during the so-called ringdown phase—i.e., the post-merger dynamics of a binary black hole system~\cite{Nollert99_qnmReview,Kokkotas:1999bd,berti,zhidenko}.
The ringdown consists of an exponentially damped oscillatory signal characterized by the QNM frequencies, i.e.~complex numbers encoding the decay rates and oscillation time scales of the GW. As distinctive signatures of distorted black holes, QNMs encode valuable information about the fundamental properties of the underlying spacetime~\cite{Dreyer:2003bv,Berti:2005ys,Berti:2016lat}. Indeed, the dominant QNM has been confirmed in several GW observations~\cite{ligo16,LIGOScientific:2020tif,LIGOScientific:2021sio}.

From a theoretical perspective, however, earlier investigations~\cite{Aguirregabiria:1996zy, Vishveshwara:1996jgz, nollert1, nollert2} and recent analyses~\cite{rodrigo,Jaramillo:2021tmt} suggest that the QNM spectrum of certain gravitational systems may be unstable under small perturbations of the associated wave equation. In other words, tiny modifications of the black hole spacetime—triggered, for instance, by environmental effects or deviations from general relativity—may significantly alter the QNM spectrum of the system~\cite{rodrigo,Jaramillo:2021tmt,Cheung:2021bol,Cardoso:2022whc,Cardoso:2024mrw}.
Thus, understanding the stability properties of QNMs is essential for determining the optimal way to extract the information they encode. One way to study such instabilities is through an explicit deformation of the underlying wave equations governing GW dynamics. This can be achieved, for instance, by introducing small modifications to the black hole potential~\cite{Aguirregabiria:1996zy,nollert1,rodrigo,Jaramillo:2021tmt,Cheung:2021bol}. Although reasonably motivated, such deformations are \emph{ad hoc} modifications to the wave equation, offering limited physical interpretation regarding their fundamental causes~\cite{Cardoso:2024mrw,Boyanov:2024fgc}. A more robust approach relies on exploring the analytical properties of the wave operator in the original black hole spacetime.

Wave dynamics in black hole spacetimes is a dissipative phenomenon: energy is absorbed by the black hole and carried away to the infinitely distant wave zone by GWs. Hence, the QNM problem naturally falls within the realm of non-Hermitian physics~\cite{Ashida:2020dkc}, where specialized mathematical tools from the theory of nonself-adjoint operators are required~\cite{dyatlov2019mathematical,Sjostrand2019}. Among these tools, the pseudospectrum provides a formal framework for studying the stability properties of the QNM spectrum~\cite{trefethen2005spectra}. A pseudospectrum analysis in black hole pertubation theory currently relies on the so-called hyperboloidal framework~\cite{Zenginoglu:2011jz,rodrigodida,PanossoMacedo:2024nkw}. By geometrically incorporating outgoing boundary conditions into the wave equation, this formalism plays a crucial role in formulating the QNM problem as an eigenvalue problem for a nonself-adjoint operator~\cite{rodrigo}. 

By expanding on the notion of the operator’s spectrum, the pseudospectrum introduces a topographic map on the complex plane, with level sets associated with a constant parameter $\varepsilon$ determining regions in the complex plane where eigenvalues may “migrate” under perturbations of magnitude $\varepsilon$. Refs.~\cite{rodrigo,Boyanov:2023qqf,Cownden:2023dam,Boyanov:2024fgc,Besson:2024adi} stress the need for careful quantitative assessments of this property. The spectrum itself is associated with peaks in the topographic map, corresponding to $\varepsilon \to 0$. As a consequence, spectral stability is visualized by concentric contour lines with steep gradients around an eigenvalue, whereas spectral instability is identified via widely spaced contour lines with shallow gradients extending far from eigenvalues.

With the notion of the pseudospectrum incorporated into gravity~\cite{rodrigo,Jaramillo:2022kuv}, pseudospectrum analyses have identified QNM spectral instabilities in various spacetimes~\cite{rodrigo4,rodrigo2,Sarkar:2023rhp,rodrigo3,Arean:2023ejh,Cownden:2023dam,Boyanov:2023qqf,Cao:2024oud,Luo:2024dxl,Chen:2024mon,Cai:2025irl,Siqueira:2025lww}. One of the biggest challenges is assessing the detectability of these instabilities in observational signals~\cite{Jaramillo:2021tmt,Spieksma:2024voy}. In this context, black hole analogs \cite{unruh1,Barcelo:2005fc} offer further observational strategies of the QNM instabilities, potentially allowing experimental setups to directly access their effects. These analogs, realized in systems such as Bose-Einstein condensates~\cite{Garay:1999sk,barcelo}, fluids~\cite{visser,Schutzhold:2002rf} and superfluids~\cite{Volovik:1995ja}, offer controlled environments where phenomena akin to those near black hole horizons can be studied experimentally. From a broader perspective, they provide a unique platform for studying classical phenomena and quantum gravitational effects~\cite{PhysRevLett.88.110201, PhysRevA.78.021603,weinfurtner2011,euve2016,steinhauer2016,torres2017,cromb2020,steinhauer2021,braidotti2022,Anacleto:2022lnt,syu2024acousticquasiboundstatestachyonic, deluca2025tidallovenumbersanalog,Albuquerque:2025eny,Vieira:2025ljl,Cromb:2025dqu}.

Theoretical and experimental studies of QNMs in analog black holes have been performed for stationary and axisymmetric systems endowed with angular momentum. 
On the theoretical side, numerical techniques can be employed to determine the QNM spectrum~\cite{Berti2,Cardoso1,geelmuyden2022}. Experimental efforts, on the other hand, have allowed the observation of QNM oscillations in fluids~\cite{mauricio1} and superfluids~\cite{Smaniotto:2025hqm}. These investigations shed light into the perspective of analog black hole spectroscopy~\cite{Torres1}, which consists in using QNM signals to identify characteristic parameters or properties of the system. Aiming to establish a potential observational strategy for measuring QNM instability in analog gravity, this work applies the pseudospectrum formalism to rotating analog black holes, while Ref.~\cite{Malato2025} investigates the black hole analog in the ``elephant and flea'' configuration~\cite{Cheung:2021bol}.

Our work is structured as follows. In Sec.~\ref{analog}, we introduce the draining bathtub model of an analog black hole and the formalism for treating perturbations in such a background. Section~\ref{sec:hyper} derives the hyperboloidal framework for this analog black hole and summarizes the numerical methods employed for solving the QNM problem. In Sec~\ref{results}, we find the pseudospectrum of nonrotating and rotating analog black holes described by the draining bathtub model. We also introduce deterministic perturbations to the system in order to investigate the impact of external factors on the QNM spectrum. Finally, in Sec.~\ref{discussion} we conclude our work by discussing the influence of rotation on the stability of the QNM spectrum.

\section{analog black hole}
\label{analog}
Linear perturbations that propagate on an inviscid, irrotational and barotropic fluid flow have been shown to be mathematically equivalent to scalar fields that propagate in curved spacetimes~\cite{unruh1,Barcelo:2005fc}. If $\mathbf{v}$ denotes the flow velocity, the irrotational character of the flow implies $\nabla \times\mathbf{v}=0$ and, consequently, there exists a scalar field $\psi$ such that $\mathbf{v} = \nabla \psi$. analog gravity in fluids relies on the fact that velocity perturbations $\delta \mathbf{v}$, characterized by perturbations $\delta \psi$ of the velocity potential, satisfy the massless Klein--Gordon equation in a curved geometry,
\begin{equation}
    \frac{1}{\sqrt{-g}}\partial_\mu(\sqrt{-g}g^{\mu\nu}\partial_\nu \delta \psi)=0,
    \label{kleingordon}
\end{equation}
where $g_{\mu \nu}$ and $g$ are, respectively, the metric of the analog spacetime and its determinant.
The line element associated with the analog spacetime has the form
\begin{align}
    ds^2\propto-(c^2-|\mathbf{v}|^2)d\tilde t^2-2\delta_{ij}v^id\tilde x^jd\tilde t+\delta_{ij}d\tilde x^id \tilde x^j,
    \label{linegeneral}
\end{align}
 where $v^i$ are the components of $\mathbf{v}$ and $c$ is the wave velocity.

A particular analog system is the draining bathtub (DBT) \cite{visser}, characterized by the following velocity field in standard polar coordinates $(r,\tilde \phi)$,
\begin{equation}
    \mathbf{v}=\frac{-A\hat{r
    }+B\hat{\tilde \phi}}{r},
\end{equation}
where $A$ and $B$ are constant parameters. The associated line element, given by
\begin{align}
    ds^2 =  &-\left(1-\frac{A^2+B^2}{c^2 r^2}\right)c^2  d \tilde t^2 + \frac{2A}{r}drd \tilde t-  2Bd\tilde\phi d\tilde t \nonumber\\&+ dr^2+r^2d\tilde\phi^2,
    \label{dtub}
\end{align}
represents a  $(1+2)$-dimensional analog rotating black hole. 

\subsection{Coordinate systems}
To better understand the properties of the analog black hole described by the line element \eqref{dtub}, we explore alternative coordinate systems that highlight the similarities between the DBT and the Kerr black hole. 
Consider first the transformation from the laboratory coordinates $\tilde x^\mu = (\tilde t, r, \tilde \phi)$ to a new coordinate system $x^\mu = (t,r,\phi)$, defined by\footnote{Our notation for the coordinate systems $(\tilde t, \tilde \phi)$ and $ (t, \phi)$ is the opposite of that in Ref.~\cite{Berti2}.}
\begin{subequations}
\begin{align}
d\tilde t &= d t + \frac{Ar}{r^2 c^2 - A^2} dr, \\
d\tilde \phi &= d\phi + \frac{BA}{r(r^2 c^2 - A^2)} dr.
\end{align}
\end{subequations}
With this transformation, the DBT line element \eqref{dtub} is recast in a form that resembles the equatorial slice of a Kerr black hole,
\begin{align}
ds^2 = -&\left(1 - \frac{A^2 + B^2}{c^2 r^2}\right) c^2 d t^2 +\left(1 - \frac{A^2}{c^2 r^2}\right)^{-1} dr^2\nonumber \\ - & 2 B  d \phi d t + r^2 d \phi^2.
\label{analoglinelem}
\end{align}
We remark, however, that an exact analog of the Kerr equatorial slice requires fine-tuning the velocity profile of the fluid~\cite{Visser:2004zs} and is, therefore, not achievable within the DBT model. Nevertheless, an event horizon and an ergosphere can still be identified in the DBT spacetime.

The analog event horizon is the circle defined by
\beq
r = r_h = \dfrac{A}{c},
\eeq
where the flow velocity in the radial direction and the wave velocity become equal. The region $r < r_h$ corresponds to the interior of the black hole, from which waves cannot escape. The analog ergoregion of the DBT, on the other hand, is determined by the circle of radius 
\beq
r = r_{e} = \dfrac{\sqrt{A^2 + B^2}}{c},
\eeq
where the total flow velocity and the wave velocity are equal. The ergoregion can be understood as the region where perturbations are dragged along the flow direction, in such a way that external observers always perceive waves in the ergoregion to be corotating with the flow. 

We remark that the coordinates $x^\mu = (t, r, \phi)$ are analogous to the Boyer-Lindquist coordinates of the Kerr spacetime~\cite{dolan}. Since $A$ defines the radial component of the flow velocity (associated with the draining rate), it can be interpreted as the effective mass of the analog black hole. On the other hand, $B$ defines the angular momentum of the flow, which can be linked to the spin of the analog black hole. Additionally, while the spin parameter of a Kerr black hole is limited by the mass of the black hole, the rotation parameter $B$ of the DBT is unbounded.

By considering null geodesics with vanishing angular momentum, we also introduce null coordinates $x^\mu_\pm = (t_\pm, r, \varphi_\pm)$ via
\beq
t_\pm = t \mp \dfrac{r_*(r)}{c}, \quad \varphi_\pm = \phi \mp \phi_*(r),
\eeq
with the radial tortoise ($r^*$) and ``angular tortoise'' ($\phi_*$) coordinates defined, respectively, by
\begin{subequations}
\label{tortoise}
\begin{align}
 \frac{dr^*}{dr} &=\left(1-\frac{A^2}{c^2r^2}\right)^{-1}, \label{radialtortoise} \\
 \dfrac{d\phi_*}{dr} &= \dfrac{B}{r^2 c} \left(1-\frac{A^2}{c^2r^2}\right)^{-1}.
 \end{align}
\end{subequations}
With these null coordinates, the line element \eqref{analoglinelem} reads
\bea
ds^2 &=& - \left( 1 - \dfrac{A^2 + B^2}{c^2 r^2} \right) c^2 dt_\pm^2 + r^2 d\varphi_\pm^2  \nonumber\\
&& \mp 2 c \,dt_\pm dr - 2 B \,dt_\pm d\varphi_\pm.
\eea
The coordinate systems $x^\mu_+$ and $x^\mu_-$ exhibit the same features as the outgoing and ingoing Kerr coordinates, respectively.
From the null coordinates, one derives null vectors analogous to the components of the Kinnersley null tetrad for Kerr black holes. Specifically, in terms of the Boyer-Lindquist-like coordinates $x^\mu$, we define the null vectors $\ell^\mu$ and $k^\mu$ according to
\begin{subequations}
\label{eq:null_vectors}
\begin{align}
\ell^\mu &= \left( 1 - \dfrac{A^2}{c^2 r^2} \right)^{-1} \times \nn \\
&\times \left( c^{-1} \delta^\mu_t + \left( 1 - \dfrac{A^2}{c^2 r^2} \right) \delta^\mu_r + \dfrac{B}{r^2 c} \delta^\mu_\phi \right), \\
k^\mu &= \dfrac{1}{2}\left( c^{-1} \delta^\mu_t - \left( 1 - \dfrac{A^2}{c^2 r^2} \right) \delta^\mu_r + \dfrac{B}{r^2 c} \delta^\mu_\phi \right).
\end{align}
\end{subequations}

\subsection{Wave equation}
Owing to the stationarity and axisymmetry of the DBT spacetime, as explicitly manifested in  \eqref{analoglinelem}, linear perturbations satisfying the Klein--Gordon equation \eqref{kleingordon} can be decomposed into modes of frequency $\omega$ and azimuthal number $m \in \mathbb{Z}$ according to
\begin{equation} 
    \delta \psi(t, r, \phi)=\frac{\Psi_{\omega m}(r)}{\sqrt{r}}e^{i(m\phi-\omega t)}.
    \label{ansatz}
\end{equation} 
 After substituting \eqref{analoglinelem} and \eqref{ansatz} into \eqref{kleingordon}, we find that the radial function $\Psi_{\omega m}(r)$ satisfies
\begin{equation}
    \frac{d^2 \Psi_{\omega m}}{dr_*^2} +Q\Psi_{\omega m}=0,
    \label{waveq}
\end{equation}
where $r_*$ is the tortoise coordinate defined in \eqref{radialtortoise}
and $Q=Q(\omega,m,r)$ is an effective potential associated with the analog black hole, given by~\cite{Berti2}
\begin{align}
    Q=&\frac{1}{c^2}\left(\omega-\frac{Bm}{r^2}\right)^2\nonumber\\&-\left(\frac{c^2r^2-A^2}{c^2r^2}\right)\Bigg[\frac{1}{r^2}\left(m^2-\frac{1}{4}\right)+\frac{5A^2}{4r^4c^2}\Bigg].
    \label{analogpotential}
\end{align}
To simplify notation, we set $\Psi_{\omega m} = \Psi_ m$, omitting the subscript $\omega$ in the remainder of this work. 

To set up the QNM problem for the DBT spacetime, we need first to determine the asymptotic solutions of the wave equation \eqref{waveq}. Far away from the analog black hole, i.e., in the limit   $r \to \infty$ (or, equivalently, $r_* \to \infty$), the general solution has the form
\begin{equation}
    \Psi_m \rightarrow Y_m^{\rm in} e^{-i \omega r_*} + Y_m^{\rm out} e^{+i \omega r_*}, 
\end{equation}
with the coefficients $Y_m^{\rm in}$ and $Y_m^{\rm out}$ representing, respectively, the amplitudes of ingoing and outgoing waves. On the other hand, near the analog event horizon, i.e., in the limit   $r \to r_h$ (or, equivalently, $r_* \to -\infty$), the general solution of \eqref{waveq} is given by
\begin{equation}
     \Psi_m \rightarrow Z_m^{\rm in} e^{-i (\omega-Bm) r_*} + Z_m^{\rm out} e^{+i (\omega-Bm) r_*},
\end{equation}
with $Z_m^{\rm in}$ and $Z_m^{\rm out}$ corresponding, respectively, to the amplitudes of the waves that enter and exit the black hole.
The one-directional nature of the analog event horizon requires that $Z_m^{\rm out}=0$, meaning that no perturbation escapes from the black hole. Additionally, since QNMs describe perturbations generated by the system itself, we exclude the possibility of external excitations propagating toward the black hole at infinity, and hence $Y_m^{\rm in}=0$. In other words, as mentioned in the introduction, the QNM problem is characterized by boundary conditions that allow only modes that are absorbed at the event horizon and propagate away from the black hole at infinity.

The boundary conditions discussed above are satisfied only for a discrete set of complex frequencies $\omega$, called QNM frequencies. The real part of the frequency describes the oscillation rate of the mode. Typically, the imaginary part of the frequency is negative, indicating that the corresponding QNM decays with time. In this case, $|\text{Im}(\omega)|^{-1}$ represents the characteristic decay time of the mode. On the other hand, QNMs presenting $\text{Im}(\omega) > 0$ imply that the system is mode unstable, indicating the potential existence of dynamic instabilities. In such a case, the quantity $|\text{Im}(\omega)|^{-1}$ corresponds to the growth rate of the unstable mode.
It is important to remark that there are no known unstable modes for the DBT spacetime~\cite{Cardoso1}. Furthermore, since there is an infinite number of QNMs for a given choice of the azimuthal parameter $m$, one conveniently introduces the overtone index $n \in \mathbb{N}$ to classify the modes. The standard convention, which we follow here, is to order the QNMs according to the absolute value of the imaginary part of their frequencies, with the fundamental mode ($n = 0$) having the smallest imaginary part (decaying most slowly), and higher overtones decaying faster. To distinguish between different azimuthal numbers and overtones, we shall use subindices $n$ and $m$ when necessary -- e.g.,~we can differentiate the QNMs by denoting their frequencies as $\omega_{n,m}$.

\section{Hyperboloidal framework}
\label{sec:hyper}

Several methods have been employed to study the QNM spectrum of the DBT model. In particular, the WKB approximation, as presented in Ref.~\cite{Berti2}, has been applied to explore the regime of small black hole rotation, while Leaver's method~\cite{Leaver:1985ax}, discussed in Ref.~\cite{Cardoso1} for the DBT model, has enabled the numerical analysis of QNMs even in large rotation scenarios. Ref.~\cite{Dolan:2011ti}, on the other hand, investigates QNMs not only in the frequency domain, through the continued fraction method, but also in the time domain, by applying finite-difference methods. Here, we introduce a new strategy, based on the hyperboloidal formalism~\cite{Zenginoglu:2011jz,rodrigodida,PanossoMacedo:2024nkw}, to calculate QNMs in analog black holes described by the DBT model.

As discussed in the previous section, the coordinate system $x^\mu=(t,r, \phi)$ parametrizing the line element \eqref{analoglinelem} shares similarities with the Boyer-Lindquist coordinates of the Kerr spacetime. Thus, along the hypersurfaces $t=$ constant, the limits $r_* \to -\infty$ and $r_* \to +\infty$ correspond, respectively, to geometrical loci akin to the black hole bifurcation sphere ${\cal B}$ and the spatial infinity $i^o$, where the QNM eigenfunctions are known to diverge. However, the physically relevant regions for wave propagation in black hole spacetimes are the black hole event horizon $\H^+$ and the future null infinity $\scri^+$, which formalizes the notion of an infinitely far away wave zone. Hyperboloidal hypersurfaces $\tau=$ constant extend smoothly between $\H^+$ and $\scri^+$, and the framework is naturally adapted to the QNM problem. 
This framework not only provides an alternative infrastructure for computing the QNMs of the DBT metric, but also enables a more detailed investigation of the spectrum's stability. Hence, in this section we also introduce the foundational concepts of the pseudospectrum and  outline the numerical techniques employed to solve the underlying equations.

\subsection{Hyperboloidal coordinates}\label{hyper}
Following the scri-fixing technique~\cite{Zenginoglu:2011jz}, with notation from \cite{rodrigodida}, hyperboloidal coordinates $\bar x^\mu = (\tau, \varphi ,\sigma)$ in the so-called minimal gauge~\cite{rodrigodida} relate to the Boyer-Lindquist-like coordinates $x^\mu = (t, r, \phi)$ via
\begin{subequations}
\label{eq:hyp_transfo}
\begin{align}
\label{timegauge}
    c\,t&=\lambda \bigg( \tau-H(\sigma)\bigg),\\
\label{angle}
\phi&=\varphi-\bar\phi_*(\sigma),\\
 \label{compact}
    r&= \dfrac{r_h} {\sigma},
\end{align}
\end{subequations}
with $\lambda$ representing a characteristic length scale of the spacetime, to be specified later.
As in the Kerr spacetime, the ``angular tortoise'' function $\bar \phi_*(\sigma) = \phi_*(r(\sigma))$ ensures that the outgoing null vector is well defined at $\H^+$. From Eq.~\eqref{tortoise}, one obtains
\beq
    \label{k}
    \bar \phi_*(\sigma)=\frac{B}{2A}\bigg( \ln{(1-\sigma)}-\ln{(1+\sigma)}\bigg).
\eeq
The height function $H(\sigma)$, on the other hand, ensures that, along $\tau=$ constant, the surfaces $\sigma=0$ and $\sigma = 1$ correspond, respectively, to $\scri^+$ and $\H^+$. To obtain it according to the minimal gauge strategy, we use Eq.~\eqref{tortoise} to define the dimensionless tortoise radial coordinate 
\bea
x(\sigma) &=& \dfrac{r_*(r(\sigma))}{\lambda} \nn \\
 &=& \dfrac{r_h}{\lambda} \bigg(  \dfrac{1}{\sigma} + \dfrac{1}{2} \ln(1-\sigma) -  \dfrac{1}{2} \ln(1+\sigma) \bigg).
\eea
The minimal gauge height function follows by changing the sign of the term that is singular at $\sigma=0$~\cite{rodrigodida}, yielding
\begin{align}
    H(\sigma)&=\frac{r_h}{\lambda}\left(-\frac{1}{\sigma}+\frac{1}{2}\ln{
 (1-\sigma)}-\frac{1}{2}\ln{(1+\sigma)}\right).
\end{align}

For convenience, we set the length scale $\lambda = r_h$ from here on. 
Using the coordinate transformation \eqref{eq:hyp_transfo}, the line element \eqref{analoglinelem} conformally rescales into $d s^2 = \Omega^{-2} d\bar s^2$, with $\Omega = \sigma/r_h$ and 
\bea
&& d\bar s^2 = - \sigma^2 \left[ 1 - \left( \dfrac{r_e}{r_h} \right)^2 \sigma^2 \right] d\tau^2 + \left( \dfrac{2r_e}{r_h} \right)^2 d\sigma^2 + d\varphi^2 \nn   \\
&&+ 2 \left[ 1 - 2 \left( \dfrac{r_e}{r_h} \right)^2\sigma^2  \right] d\tau d\sigma  
 - 2 \dfrac{B}{A} \sigma^2 d\tau d\varphi + 4\dfrac{B}{A}d\sigma d\varphi. \nn \\
\eea
Particularly relevant to the wave equation \eqref{kleingordon}, the inverse metric assumes a rather simple form, with components
\begin{subequations} \label{inverse_hyp_metric}
\begin{align}
&\bar g^{\tau \tau}  = - 4 =: - w(\sigma), \\ 
&\bar g^{\sigma \sigma} = \sigma^2(1-\sigma^2) =: p(\sigma), \\
&\bar g^{\tau \sigma} = 1-2\sigma^2 =: \gamma(\sigma), \\ 
&\bar g^{\sigma \varphi}  = -\dfrac{B}{A}\sigma^2=: \alpha(\sigma),\\
& \bar g^{\varphi \varphi} = 1, \quad  \bar g^{\tau \varphi}  = -\dfrac{2B}{A}.
\end{align}
\end{subequations}

We verify that the coordinates $\bar x^\mu$ are indeed hyperboloidal. First, the vector that is normal to the surfaces $\tau=$ constant is timelike outside the black hole since 
\beq
|| \bar \nabla_a \tau|| = \bar g^{\tau \tau} <0, \quad \text{for} \quad \sigma \in [0,1].
\eeq
Hence, the hyperboloidal surfaces are spacelike in the exterior black hole region. 
Second, we consider the conformal null vectors $\bar \ell^{\bar \mu}$ and $\bar k^{\bar \mu}$, defined by 
\beq
\bar \ell^{\bar \mu} = \dfrac{\zeta}{\Omega} \ell^{\bar \mu}, \quad \bar k^{\bar \mu} = \dfrac{1}{\zeta \Omega} k^{\bar \mu}, 
\eeq
where $\zeta = (1-\sigma^2)/\sigma$ and the components of the null vectors $\ell^{\bar \mu}$ and $k^{\bar \mu}$, given in \eqref{eq:null_vectors}, are expressed in hyperboloidal coordinates $\bar x^\mu$. Although the factor $\Omega^{-1}$ arises naturally from the conformal transformation, the boost parameter $\zeta$ ensures the regularity of the conformal null vectors for $\sigma \in [0,1]$. Their components read explicitly
\beq
\bar \ell^{\bar \mu} = 2 \delta^{\bar \mu}_{\tau} - (1-\sigma^2) \delta^{\bar \mu}_{\sigma} + \dfrac{2 B}{A} \delta^{\bar \mu}_{\varphi}, \quad \bar k^{\bar \mu} =  \delta^{\bar \mu}_{\tau} + \dfrac{\sigma}{2}\delta^{\bar \mu}_{\sigma}.
\eeq
The surface $\sigma = 0$ is generated by the ingoing null vector $\bar k^{\bar \mu} = \delta^{\bar \mu}_{\tau}$, whereas the surface $\sigma =1$ is generated by the outgoing null vector $\bar \ell^{\bar \mu} \propto \delta^{\bar m}_{\tau}$. These expressions confirm that the $\tau =$ constant hypersurfaces intersect future null infinity at $\sigma = 0$ and the black hole at $\sigma = 1$, as expected.

The hyperboloidal representation of the DBT spacetime offers a simplified approach to investigating the effects of rotation on the spectrum stability of analog black holes.
From Eq.~\eqref{inverse_hyp_metric}, we observe that the functions $p$, $\gamma$ and $w$ are defined in the same way that they are defined in spherically symmetric spacetimes~\cite{rodrigodida}, i.e.
\begin{subequations} \label{eq:def_hyp_func}
\begin{align}
p(\sigma)&:= -\dfrac{1}{x'(\sigma)},\\
\gamma(\sigma) &:= p(\sigma)H'(\sigma),\\  
w(\sigma)&:= \dfrac{1-\gamma(\sigma)^2}{p(\sigma)}.
\end{align}
\end{subequations}
However, the transformation of the angular coordinate \eqref{eq:hyp_transfo} introduces the extra function
\beq
\label{eq:def_hyp_alpha}
\alpha(\sigma):= p(\sigma)\bar{\phi}_*'(\sigma).
\eeq

\subsection{QNM eigenvalue problem}
With the help of the hyperboloidal formalism introduced in the previous section, we reformulate the QNM problem discussed in Sec.~\ref{analog} to re-cast it as an eigenvalue problem associated with a nonself-adjoint operator. The coordinate transformation \eqref{eq:hyp_transfo} applied to the ansatz \eqref{ansatz} induces a rescaling in the frequency domain function of the form
\begin{subequations}
\label{eq:hyp_freq_transfo}
\begin{align}
\Psi_m(r(\sigma)) &= \mathcal{Z}_m(\sigma) \bar \Psi_m(\sigma), \\ \mathcal{Z}_m(\sigma) &= e^{s H(\sigma) + i m \bar \phi_*(\sigma)},
\end{align}
\end{subequations}
where $s = - i \lambda \omega$ is a dimensionless frequency parameter. 
As expected for the hyperboloidal framework in the minimal gauge, the rescaling term $Z_m(\sigma)$ in Eq.~\eqref{eq:hyp_freq_transfo} coincides with the prefactor normally introduced when approaching the QNM problem via Leaver's method~\cite{Leaver:1985ax,berti}. However, instead of Taylor expanding the regular function $\bar \Psi_m(\sigma)$ around the horizon, and solving the resulting equation with the help of a continued fraction algorithm, we employ a Chebyshev spectral approximation to describe $\bar \Psi_m(\sigma)$, as detailed in the upcoming Sec.~\ref{sec:NumMeth}.

Substituting \eqref{eq:hyp_freq_transfo} into the wave equation \eqref{waveq}, and introducing the auxiliary field $\bar \Phi_m = s \bar \Psi_m$, leads to a reformulation of the QNM problem as an eigenvalue problem for
\beq
\label{eq:QNM_Eigenvalue}
{\rm L}\, \boldsymbol{u}_m = s \, \boldsymbol{u}_m,
\eeq
with 
\beq
   {\rm L} = \begin{pmatrix} 0 & 1 \\ {\rm L}_1 & {\rm L}_2 \end{pmatrix}, \quad \boldsymbol{u}_m  = \begin{pmatrix} \bar \Psi_m \\ \bar \Phi_m
     \end{pmatrix},
    \label{operatorL}
\eeq
and
\begin{subequations}
\label{eq:L1L2}
\begin{align}
\label{l1}
   & {\rm L}_1 = \frac{1}{w(\sigma)} \Bigg[ \partial_\sigma (p(\sigma) \partial_\sigma) - q_m(\sigma) + 2 i m \alpha(\sigma) \partial_\sigma \Bigg], \\
    \label{l2}
 &   {\rm L}_2 = \frac{1}{w(\sigma)} \Bigg[ 2 \gamma(\sigma) \partial_\sigma + \gamma'(\sigma)+\frac{2 i m\alpha(\sigma)}{p(\sigma)} \left( \gamma(\sigma) +1 \right)\Bigg].
\end{align}
\end{subequations}
The function $q_m$ is the effective potential, given by
\begin{equation}
    q_m(\sigma)= \left[\frac{5\sigma^2+4m^2-1}{4} + \dfrac{2 i m B\sigma}{A}\right].
    \label{effectivepotential}
\end{equation}
The hyperboloidal functions $w$, $p$, $\gamma$ and $\alpha$ were defined in terms of the inverse conformal metric in Eq.~\eqref{inverse_hyp_metric}, or equivalently in Eqs.~\eqref{eq:def_hyp_func}-\eqref{eq:def_hyp_alpha}. We observe that the last term in Eq.~\eqref{l2} simplifies to
\beq
\frac{2 i m\alpha(\sigma)}{p(\sigma)} \left( \gamma(\sigma) +1 \right) = -\dfrac{4 i B m}{A},
\eeq
remaining regular at the boundaries, where $p(0)=p(1)=0$. 

The QNMs now follow directly as the eigenvalues $s_n$ of the operator ${\rm L}$. In particular, a comparison between the expressions for ${\rm L}_1$ and ${\rm L}_2$ in \eqref{eq:L1L2} with those used in the case of spherically symmetric spacetimes~\cite{rodrigodida} highlights the presence of the extra terms $i m \alpha(\sigma) \sim i m B/A$ due to the rotational nature of the DBT spacetime. This additional term breaks the symmetry between the eigenvalues with ${\rm Im}(s)>0$ and ${\rm Im}(s)<0$. Based on the propagation of the surfaces of constant phase, one can then distinguish prograde and retrograde modes through the following:
\begin{subequations}
\begin{align}
&\text{Prograde modes:} \ 
{\rm sign} \left[ {\rm Im} \, \left(s_{n,m}\right) \right] = - {\rm sign} [m],
\\
&\text{Retrograde modes:}  \
{\rm sign} \left[ {\rm Im} \, \left(s_{n,m}\right) \right] = {\rm sign} [m].
\end{align}
\label{prograde}
\end{subequations}
We note that the eigenvalues also possess the property
\beq
\label{eq:Mirror_modes}
s_{n,m} = s^\ast_{n,-m},
\eeq
where the symbol $*$ denotes the complex conjugate operation.

\subsection{Pseudospectrum and condition number}
The pseudospectrum associated with the operator ${\rm L}$ offers a robust tool to investigate the stability properties of the QNM spectrum in the DBT spacetime \cite{rodrigo}. The $\varepsilon$-pseudospectrum $\varsigma_{\varepsilon}(A)$ is defined by \cite{trefethen2005spectra}
\begin{equation}
   \hspace{-0.20cm}  \varsigma_{\varepsilon}({\rm L})=\{s\in\mathbb{C}:||{\cal R}_{{\rm L}}(s)||>1/\varepsilon\},
    \label{pseudospectrum}
\end{equation}
where ${\cal R}_{{\rm L}}(s) = (s \rm Id-{\rm L})^{-1}$ is the resolvent of the operator and $\varepsilon>0$ is a fixed parameter. 
For $\varepsilon=0$, one recovers the definition of the QNM spectrum $\varsigma_{0}({\rm L}) = \varsigma({\rm L})$. For $\varepsilon\neq 0$, the $\varepsilon$-pseudospectra level sets enclose regions in the complex plane where eigenvalues may shift under perturbations of magnitude $\varepsilon$ to the operator $L$. Thus, the structure of the contour lines provides insight into the spectral stability of the operator: concentric contour lines with steep gradients around eigenvalues indicate spectral stability, whereas widely spaced contour lines with shallow gradients extending far from eigenvalues are indicative of spectral instability.

The pseudospectrum's definition \eqref{pseudospectrum} requires the introduction of an appropriate norm to properly quantify the intensity of the parameter $\varepsilon$. A natural choice is the energy associated with the propagating scalar field~\cite{rodrigo, Gasperín}. For the system under consideration, the energy on a slice $\tau=$ constant, described by coordinates $\bar x^i = (\sigma, \varphi)$, is given by
\beq
E = \int\limits_{{\tau={\rm cont.}}} J^{\bar \mu} n_{\bar \mu} \, {\rm dVol}, \quad J^{\bar \mu} = - T^{\bar \mu \bar \nu}t_{\bar \nu},
\eeq
with $t^{\bar \mu} = (\partial_t)^{\bar \mu}$ the timelike Killing vector of the DBT spacetime, $n_{\bar \mu}$ the unit normal vector to the hypersurface $\tau =$ constant, and ${\rm dVol} = \sqrt{h} d\sigma d\varphi$ the volume element associated with the hypersurface $\tau =$ constant, with induced metric $h_{\bar i, \bar j}$.  The flux $J^{\bar \mu}$ results from the scalar field stress-energy momentum tensor
\beq
T_{\bar \mu \bar \nu} = \nabla_{\bar \mu}(\delta \psi) \nabla_{\bar \nu}(\delta \psi) - \dfrac{g_{\bar \mu \bar \nu}}{2}|| \nabla(\delta \psi) ||^2. 
\eeq
A decomposition akin to Eq.~\eqref{ansatz}, performed directly in the hyperboloidal slices, produces
\beq
\delta \psi(\tau, \sigma, \varphi) = \sqrt{\Omega(\sigma)} \sum_{m=-\infty}^{\infty} e^{s\tau} \bar \Psi_m(\sigma) e^{i m \varphi},
\eeq
yielding
\begin{align}
E &\propto \sum_{m=-\infty}^{\infty} E_m,
\end{align}
with
\begin{align}
\label{eq:E_m}
E_m &= \dfrac{1}{2}\int_0^1 d\sigma \bigg[ w(\sigma)|\Phi_m|^2 \nonumber\\
&+p(\sigma)|\partial_\sigma\Psi_m|^2+q^o_m(\sigma) |\Psi_m|^2\nonumber  
\\
& - im\alpha(\sigma)\left(\Psi_m^\ast\partial_\sigma\Psi_m-\Psi_m\partial_\sigma\Psi_m^\ast\right) \bigg],
\end{align}
and $q_m^o(\sigma) =\left.q_{m}(\sigma)\right|_{B\to 0}$ the static effective potential derived from Eq.~\eqref{effectivepotential}. 

From \eqref{eq:E_m} we define the energy norm as
\beq
\label{eq:energy_norm}
||{\boldsymbol u}_m||_{_E} =  E_m
\eeq
and the energy product between two state vectors ${\boldsymbol u}_{{}_m^1}$ and ${\boldsymbol u}_{{}_m^2}$ as
\begin{align}
   \braket{{\boldsymbol u}_{{}_m^2}|{\boldsymbol u}_{{}_m^1}}_{_{E}}&= \dfrac{1}{2}\int_0^1\Bigg[w(\sigma)\Phi_{{}_m^2}^\ast\Phi_{{}_m^1}\nonumber\\&+p(\sigma)\partial_\sigma\Psi_{{}_m^2}^\ast\partial_\sigma\Psi_{{}_m^1}+q_m^o(\sigma) \Psi_{{}_m^2}^\ast\Psi_{{}_m^1}\nonumber\\&+im\alpha(\sigma)\left(\Psi_{{}_m^2}^\ast\partial_\sigma\Psi_{{}_m^1}-\Psi_{{}_m^1}\partial_\sigma\Psi_{{}_m^2}^\ast\right)\Bigg].
    \label{innerproduct}
\end{align}

The energy product \eqref{innerproduct} is also employed to calculate the condition number
\beq
\kappa_{n,m} = \dfrac{ ||{\boldsymbol v}_{n,m}||_{_E} \, ||{\boldsymbol u}_{n,m}||_{_E}   }{\left| \braket{{\boldsymbol v}_{n,m} |{\boldsymbol u}_{n,m}}_{_{E}}\right|},
\label{eq:cond_number}
\eeq
with ${\boldsymbol u}_{n,m}$ the (right) QNM eigenfunction of the operator ${\rm L}$ associated with the QNM frequency $s_{n,m}$, and ${\boldsymbol v}_{n,m}$ the corresponding (left) eigenfunction of the operator ${\rm L}^\dagger$.
In addition to the pseudospectra, the condition number provides another tool for diagnosing the instability of a given QNM $s_{n,m}$. In particular, $\kappa_{n,m} \approx 1$ indicates spectral stability, whereas $\kappa_{n,m} \gg 1$ signalizes spectral instability.

In the next section, we summarize the numerical techniques employed to calculate the QNM spectra, the operator's pseudospectrum, and the eigenvalue condition numbers.

\subsection{Numerical methods}
\label{sec:NumMeth}
The numerical discretization of the operator \eqref{operatorL} is carried out through a spectral scheme based on a Chebyshev collocation point method~\cite{spectralmethods}. To achieve this, we fix a parameter $N\in{\mathbb N}$ and use the Chebyshev-Lobatto grid points to discretize the domain $\sigma\in[0,1]$ into
\beq
\label{eq:sigma_grid}
\sigma_i = \dfrac{1+x_i}{2}, \quad x_i = \cos\left(\dfrac{i \pi}{N} \right),\quad i=0\cdots N.
\eeq
We also assume an approximation of the unknown function $\bar \Psi_m(\sigma)$ in terms of the Chebyshev polynomials of first kind, $T_k(x) = \arccos(k \cos(x))$, via
\beq
\Psi_m(\sigma(x)) \approx \sum_{k=0}^N c_k T_k(x),
\eeq
and impose that the approximation is exact at the grid points $\sigma_i$. This scheme allows one to derive a discrete representation for the derivative and integration operators, namely
\beq
    \partial_x \rightarrow \hat D_{x}, \quad \int_{-1}^{1}dx \rightarrow \hat {\cal G}_{x}.
\eeq
For details about the components of the matrices $\hat D_{x}$ and $\hat {\cal G}_{x}$, we refer to the appendix of \cite{rodrigo}. With the help of Eq.~\eqref{eq:sigma_grid}, we obtain the relavant operators for the domain $\sigma\in[0,1]$ as  
\beq
    \label{eq:disc_derv}
    \hat D_{\sigma} = 2  \hat D_{x}, \quad \hat{\cal G}_{\sigma} = \dfrac{1}{2} \hat{\cal G}_{x}. 
    \eeq
    
Evaluating the state vector ${\boldsymbol u}_m$ at the grid points \eqref{eq:sigma_grid} produces its discrete representation $\vec {\boldsymbol u}_m$, whose length is $n_{\rm total} = 2 (N+1)$ and whose components are
\beq
\left( \vec{{\boldsymbol u}}_m \right)_i = {\boldsymbol u}_m(\sigma_i) = \begin{pmatrix} \bar \Psi_m(\sigma_i) \\ \bar \Phi_m(\sigma_i)
     \end{pmatrix}.
\eeq
Then, the discrete representation of the operator ${\rm L}$, which acts on $\vec {\boldsymbol u}_m$ as in Eq.~\eqref{eq:QNM_Eigenvalue}, reads
\beq
\hat{L}= \left( 
    \begin{array}{cc}
    \mathbb{0} & \mathbb{1} \\
    \hat{L}_1 & \hat{L}_2
    \end{array}
    \right),
\eeq
where
    \begin{subequations}
    \bea
        \hat{L}_1 &=& \vec{w}^{-1}\circ \left( \vec{p}\circ \hat D^2_{\sigma} +\vec{p'}\circ \hat D_{\sigma} - \vec{q}_{\ell}\circ \mathbb{1}  + 2 i m \vec\alpha\circ \hat D_{\sigma}\right), \nn \\
    \\
    \hat{L}_2 &=& \vec{w}^{-1}\circ \left( 2\vec{\gamma}\circ \hat D_{\sigma} +\vec{\gamma'}\circ \mathbb{1} -\dfrac{4 i B m}{A}\mathbb{1}  \right).
    \eea
    \end{subequations}
Similarly, the energy norm \eqref{eq:energy_norm} assumes the discrete form
    \begin{subequations}
     \bea
	 && ||\vec {\boldsymbol u}_m||^2_{_{E}}  = \vec{{\boldsymbol u}}_m^\ast \cdot \hat G \cdot \vec {\boldsymbol u}_m,  \\
     && \hat G = \left(
        \begin{array}{c|c}
      \hat D_{\sigma}^\intercal \cdot ( \vec p  \circ \hat{\cal G}_{\sigma}  ) \cdot \hat D_{\sigma} +  \vec q \circ \hat{\cal G}_{\sigma}\\+i m \left[\hat D_{\sigma}^\intercal\cdot (\vec{\alpha} \circ \hat{\cal G}_{\sigma} )-(\vec{\alpha} \circ \hat{\cal G}_{\sigma} )\cdot\hat D_{\sigma}\right]     & \mathbb{0} \\
      \hline
           \mathbb{0}  & \vec w \circ \hat{\cal G}_{\sigma}
        \end{array}
        \right). \nn \\
     \eea
    \label{energyintegration}
     \end{subequations}
We note that $\hat G$ is also employed to calculate the products
\beq
\braket{ \vec {\boldsymbol v}_m | \vec {\boldsymbol u}_m}_{_E} = \vec {\boldsymbol v}_m^\ast \cdot G \cdot \vec {\boldsymbol u}_m
\eeq
for the condition number \eqref{eq:cond_number}.

In all above expressions, we have employed the following definitions. For a given function $f(\sigma)$, the vector $\vec f$ has components $(\vec f)_i = f(\sigma_i)$, and we abuse the notation to define $\vec f^{-1}$ via $(\vec f^{-1})_i = (f_i)^{-1}$. The circle ($\circ$) operator denotes the element-wise (Hadamard) products, defined for a vector $\vec f$ and matrix $\hat M$ by
    \beq
    \left(\vec{f}\circ \hat M\right)_{ij} = f_i \, M_{ij}.
    \eeq
Additionally, $\mathbb{0}$ and $\mathbb{1}$ are, respectively, the null and identity matrices of size $(N+1)\times (N+1)$.
       
Within this discretization scheme, the $\varepsilon$-pseudoespectrum $\varsigma^{\varepsilon}(\hat L)$ is determined by
\begin{equation}
		\varsigma_{\varepsilon} (\hat L) = \{ s \in \mathbb{C}: {\mathfrak s}^{\text{min}} \left( \hat A \right) < \varepsilon \}, \quad \hat A = \lambda \mathbb{I} - \hat L,
        \label{discretizepseudo}
	\end{equation}
    where $\mathbb{I}$ is the identity matrix of dimension $n_{\rm total}\times n_{\rm total}$,
	and ${\mathfrak s}^{\text{min}}$ is the smallest of the generalized singular values of the matrix $\hat A$, namely
	\begin{equation}
		{\mathfrak s}^{\text{min}}(\hat A) = \text{min} \left\{\sqrt{s} : s \in \varsigma \left(\hat A^{\dagger}\hat A\right)\right\},
	\end{equation}
    with $\varsigma \left( \hat A^{\dagger}\hat A\right)$ the spectrum of the operator $\hat A^{\dagger}\hat A$. To calculate ${\mathfrak s}^{\text{min}}$ with the above expression, one requires the discrete adjoint $\hat{L}^{\dagger}$ of the operator $\hat L$ with respect to the energy scalar product, given by
	\begin{equation}
		\hat L^{\dagger} = \left(\hat G\right)^{-1} \cdot \left(\hat L^{*}\right)^t \cdot \hat G.
	\end{equation}

\section{Results}
\label{results}

Using the framework developed in Sec.~\ref{sec:hyper}, we now examine the impact of rotation on the QNM spectrum of a rotating analog black hole described by the DBT metric, with a particular focus on its stability properties. In Sec.~\ref{qnmpseudo}, we compute the QNM spectrum within the hyperboloidal framework and analyze its behavior as the rotation parameter increases. Subsequently, in Sec.~\ref{pseudospectrumrotating}, we evaluate the associated pseudospectra for different DBT configurations, highlighting the influence of rotation on their overall structure. Finally, in Sec.~\ref{instability}, we investigate the effect of deterministic perturbations on the effective potential of the system, assessing their impact on the analog black hole across different rotation regimes.

We focus on the azimuthal number $m=1$, as the corresponding QNMs decay more slowly than those with higher $m$ values, making them more likely to be detected. The truncation parameter $N$ associated with the Chebyshev method is chosen according to the specific requirements of each analysis.  In general, the procedure followed throughout is as follows: to ensure numerical convergence, we require that the relative error between successive computations of the third overtone ($n=3$), at $N$ and $N+50$, remains below a prescribed threshold. For the QNM calculations in Sec.~\ref{qnmpseudo}, $N = 300$ ensures that the relative error is always below $10^{-3}$. This resolution is then fixed for the pseudospectrum analysis and the condition number calculations in Sec.~\ref{pseudospectrumrotating}. For the computation of perturbed QNMs in Sec.~\ref{instability}, we observed an increased sensitivity with respect to $N$. In particular,  stationary flows ($B=0$) require a truncation up to $N=1000$ to guarantee a relative error of order $\sim 10^{-2}$ in the calculation of the third overtone. For $B>0$, the choice $N=500$ is sufficient to achieve a relative error of $\sim 10^{-3}$.

\subsection{Quasinormal modes}
\label{qnmpseudo}

\begin{figure}[t!]
\centering
\includegraphics[scale=0.9]{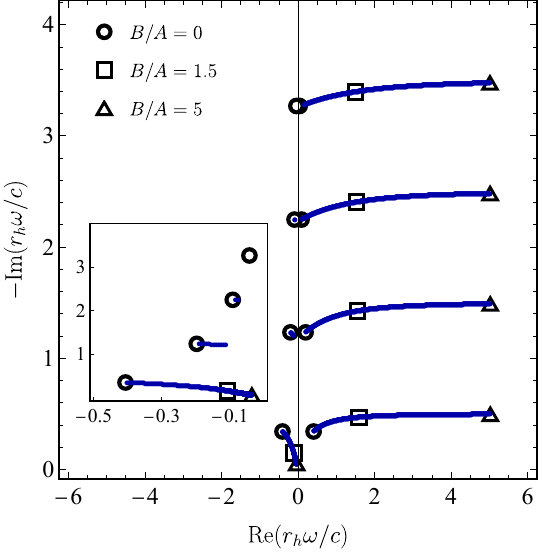}
    \caption{The first four prograde ($\rm Re(\omega) > 0$) and retrograde ($\rm Re(\omega) < 0$) QNMs of the $m=1$ perturbations in the DBT model. The inset provides an enlarged view of the retrograde modes. Along each curve, the parameter $B/A$ varies from $0$ to $5$. For the prograde modes, both the real and imaginary parts increase with $B/A$, with the real parts coalescing to a common value as rotation increases. Conversely, the retrograde modes exhibit a decrease in both the real and imaginary parts as $B/A$ grows.
    We note that, for $n=1,2,3$, we are no longer able to track the retrograde QNMs when the rotation parameter becomes large enough and the associated frequencies approach the imaginary axis.} 
    \label{fig1}
\end{figure}

With the hyperboloidal framework introducing a new approach to calculating QNMs in analog spacetimes, we begin by benchmarking our numerical strategy, based on a Chebyshev discretization scheme (cf.~Sec.~\ref{sec:NumMeth}), against results obtained through Leaver's continued fraction method~\cite{Cardoso1}. The advantage of solving the QNM problem as the direct eigenvalue problem, given by Eq.~\eqref{eq:QNM_Eigenvalue}, is that we obtain all QNMs simultaneously (within numerical roundoff error), without relying on a root-finding algorithm that requires a fine-tuned initial seed for a specific $\omega$ value. Indeed, Fig.~\ref{fig1} tracks all QNMs up to overtone $n=3$ as a function of the rotation parameter in the range $B/A \in [0,5]$.

We first concentrate on the nonrotating case $B/A=0$, whose QNMs correspond to circles in Fig.~\ref{fig1}. Due to the absence of rotation, the QNMs are symmetric with respect to the imaginary axis, with both $\omega_{m,n}$ and $-\omega_{m,n}^\ast$ belonging to the QNM spectrum. The tendency of QNMs to approach the imaginary axis for higher overtones is also observed, which is consistent with previous findings \cite{Cardoso1}. This behavior presents an obstacle to identifying overtones beyond $n=3$ for nonrotating and slowly rotating analog black holes. Notably, this limitation is not specific to our numerical scheme, as it also appears in Leaver's method.

In our approach, the difficulty in resolving modes near the imaginary axis stems from the presence of spurious, nonconverging eigenvalues associated with the discrete operator $\hat{L}$. As observed in other contexts, QNMs form only a subset of the eigenvalues of this discrete operator~\cite{rodrigo}. For a given resolution $N$, additional spurious modes appear, which fail to converge to fixed values as $N$ increases. In Schwarzschild spacetime, some of these spurious modes are distributed precisely along the imaginary axis and are interpreted as a discrete representation of the associated branch cut~\cite{Leaver:1986gd,rodrigo}. 
Similarly, in the DBT model, spurious modes are found clustered within a small region around the imaginary axis. Consequently, resolving actual QNMs with values close to the imaginary axis becomes increasingly challenging as the overtone number grows. To the best of our knowledge, no systematic analytical study exists on the asymptotic behavior of QNMs for the DBT model in the large-$n$ regime. 

As the rotation parameter $B/A$ increases from zero, the symmetry between $\omega_{m,n}$ and $-\omega_{m,n}^\ast$ is broken, as shown in Fig.~\ref{fig1}. Instead, relation \eqref{eq:Mirror_modes} introduces a symmetry between the modes $\omega_{m,n}$ and $-\omega_{-m,n}^\ast$.   
Consequently, as the rotation parameter is varied, the trajectories of modes with $\rm Re (\omega) > 0$ differ significantly from those with $\rm Re (\omega) < 0$. In this paper, where we consider a positive azimuthal number $m=1$, they correspond to prograde and retrograde modes, respectively. As reported in Ref.~\cite{Cardoso1}, the real part of the prograde modes ($\rm Re (\omega) > 0$) begins to coalesce as rotation increases, independent of the overtone, and continues to grow linearly with $B/A$. Meanwhile, the imaginary part remains distinct across overtones, with higher overtones exhibiting larger decay rates. In other words, as rotation increases, prograde modes oscillate at higher frequencies and decay more rapidly, eventually tending to oscillate at the same frequency in the high-rotation regime. 
In contrast, retrograde modes ($\rm Re (\omega) < 0$) exhibit a decrease in both real and imaginary parts as rotation increases, leading to slower oscillations and lower decay rates. As also observed in Ref.~\cite{Cardoso1}, for each overtone $n$, there exists a critical value of $B/A$ at which the QNM reaches the imaginary axis, i.e., $\rm Re(\omega) = 0$. Beyond this threshold, the overtone is no longer identifiable. In the limit $B/A \rightarrow \infty$, the only remaining retrograde QNM corresponds to the fundamental mode ($n=0$), with $\omega \rightarrow 0$.

\begin{figure*}[!htpb]
		\centering
		\includegraphics[width = 1 \linewidth]{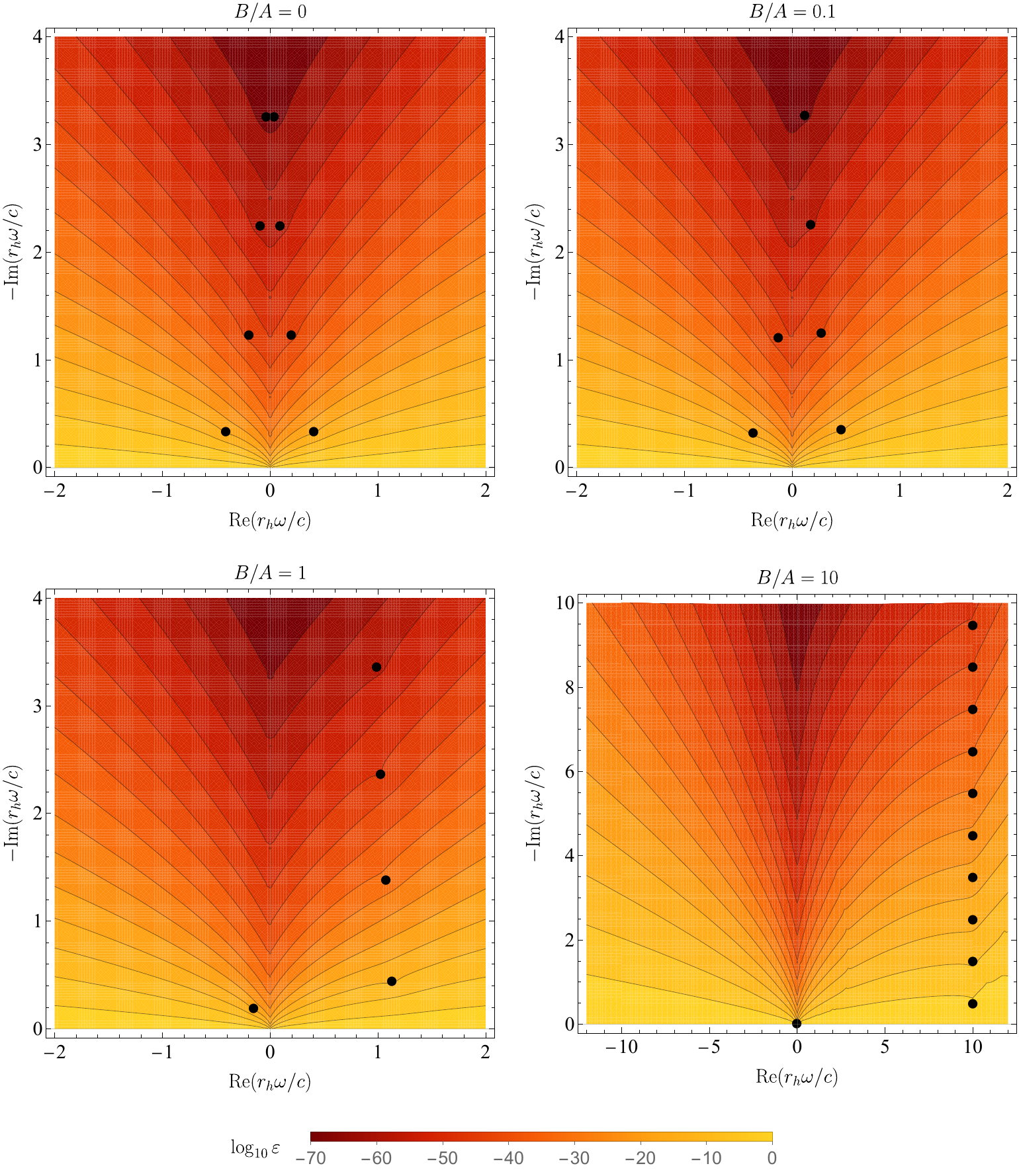}
		\caption{The pseudospectrum of the $m=1$ scalar perturbations in the DBT model is shown for different values of the rotation parameter $B$. The top panels correspond to $B/A = 0$ (left) and $B/A = 0.1$ (right), while the bottom panels represent $B/A = 1$ (left) and $B/A = 10$ (right). In each panel, black circles represent the QNMs. For $B/A = 0, 0.1$, and $1$, the QNMs include the fundamental mode and the first three overtones. In contrast, the case $B/A = 10$ contains a higher number of QNMs, extending from the fundamental mode up to the ninth overtone. The color scale in the background of each panel represents $\log_{10} \varepsilon$, and the black contour lines denote level sets of constant $\varepsilon$. These contour lines indicate spectral instability, as they spread across the complex plane. The pseudospectrum structure is asymmetric in rotating scenarios, and the QNMs are dragged through it as rotation increases. The retrograde modes $(\mathrm{Re}(\omega) < 0)$ shift leftward, asymptotically approaching zero with increasing rotation, whereas the prograde modes $(\mathrm{Re}(\omega) > 0)$ shift rightward. Notably, for $B/A = 10$, the lower prograde overtones, namely $n=1$ and $n=2$, begin to migrate to the region occupied by the fundamental mode, suggesting that, as rotation increases, overtones tend to occupy areas of potentially enhanced stability in the pseudospectrum.}
		\label{fig2} 	
	\end{figure*}

\subsection{Stability analysis: Pseudospectra and condition number}
\label{pseudospectrumrotating}
After benchmarking the new framework for calculating the QNM spectra, we proceed to analyze its stability using the pseudospectrum and the condition number. Using the definition of the pseudospectrum given in Eq.~\eqref{discretizepseudo}, along with the discretized energy norm from Eq.~\eqref{energyintegration}, we obtain the pseudospectrum of the DBT model for scalar perturbations with azimuthal number $m=1$. Figure~\eqref{fig2} presents the results for different rotation regimes, with QNMs indicated by black dots. 

The top left panel displays the results for the nonrotating case ($B/A=0$), where we observe that the contour lines of the pseudospectrum extend widely across the complex plane. As discussed previously, this behavior signals spectral instability, as the boundaries of the $\varepsilon$-pseudospectrum define the potential migration zones of the QNMs. Additionally, notice that the pseudospectrum in this case is symmetric, following the overtone symmetry discussed in Sec.~\ref{qnmpseudo} for the nonrotating regime. As rotation increases and this symmetry is broken, the overall structure of the pseudospectrum becomes asymmetric.
This asymmetry trend persists for both small rotation values ($B/A = 0.1$, top-right panel) and moderate values ($B/A = 1$, bottom-left panel), and increases in the high-rotation regime ($B/A = 10$, bottom-right panel). As observed in the previous section, increasing $B/A$ results in all prograde modes acquiring approximately the same value of ${\rm Re}\left(r_h \omega/c \right)$. 

\begin{figure}[t!]
\centering
\includegraphics[width =\columnwidth ]{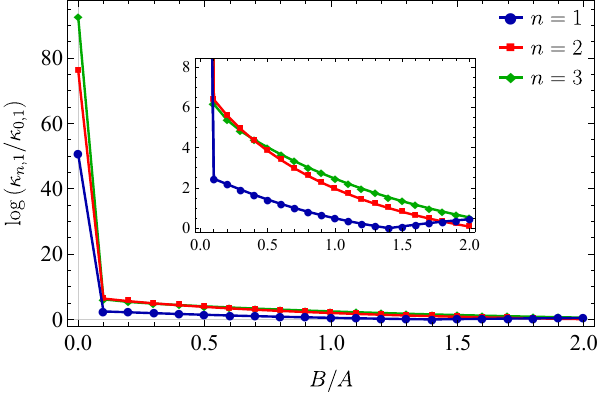}
    \caption{The relative ratio of the condition number between the overtones and the fundamental mode as a function of rotation. The inset provides a detailed view of the region \( B/A \in [0.1, 2] \). In the nonrotating case (\( B/A = 0 \)), this ratio is significantly high, indicating that the overtones do not share the stability properties of the fundamental mode under small perturbations. As rotation increases, from $B/A=0$ to $B/A=0.1$, the ratio exhibits a sharp decline. As rotation continues to increase, the condition numbers of the prograde overtones approach that of the fundamental mode, which potentially leads to similar stability properties. For \( B/A > 2 \), numerical noise introduces significant variations on the relative ratio, complicating further analysis.}

    \label{fig:cond_numb}
\end{figure}

It is worth remarking that, as rotation increases, the prograde modes migrate across different regions of the $\varepsilon$-pseudospectrum. As a consequence, overtones, one by one, seem to approach the same region occupied by the fundamental mode. This behavior is explicitly observed in Fig.~\ref{fig2} for the first overtone when $B/A=10$, with a noticeable trend emerging for higher overtones.
We recall that for the Schwarzschild spacetime, Ref.~\cite{rodrigo} reports that the fundamental mode is stable under ultraviolet perturbations, whereas the overtones are more susceptible to spectral destabilization. With the lower overtones in the DBT model eventually occupying the same $\varepsilon$-pseudospectrum region as the fundamental mode, it is worth analyzing whether these modes have enhanced stability properties as rotation increases.
To this end, we use the condition number $\kappa_{n,m}$, cf.~\eqref{eq:cond_number}, as a complementary tool to assess the stability of the QNM spectrum. Fig.~\ref{fig:cond_numb} shows the dependence of the relative ratio $\kappa_{n,1}/\kappa_{0,1}$ between the condition number of a given prograde overtone $n$ to that of the fundamental mode $n=0$, as a function of $B/A$.

Examining the interval $B/A \in [0, 0.1]$ in Fig.~\ref{fig:cond_numb}, we observe that the introduction of rotation leads to a sharp decrease in the relative ratio of the condition numbers for all overtones. This suggests that even a modest rotation parameter drives the condition numbers of the overtones toward that of the fundamental mode.
A clear trend emerges as rotation continues to increase. We observe that the rate of decay of the relative ratio is different depending on the overtone, with $\kappa_{1,1}$ approaching $\kappa_{0,1}$ more rapidly than $\kappa_{2,1}$ and $\kappa_{3,1}$. Given the fact that the fundamental mode is stable under ultraviolet perturbations, this result strongly suggests that rotation acts to stabilize the overtones. For $B/A > 2$, however, numerical noise limits a more conclusive statement, reflecting known constraints in the convergence properties of quantities derived from the discretized adjoint operator~\cite{rodrigo,Boyanov:2023qqf,Cownden:2023dam,Boyanov:2024fgc,Besson:2024adi}. Despite these numerical challenges in the high-rotation regime $B/A > 2$, both the pseudospectrum and the general trend associated with the condition number strongly suggest that the rotation parameter $B/A$ plays a crucial role in enhancing the stability of the overtones.

\subsection{Spectrum instability}
\label{instability}

\begin{figure*}[!t]
		\centering
		\includegraphics[width = 1 \linewidth]{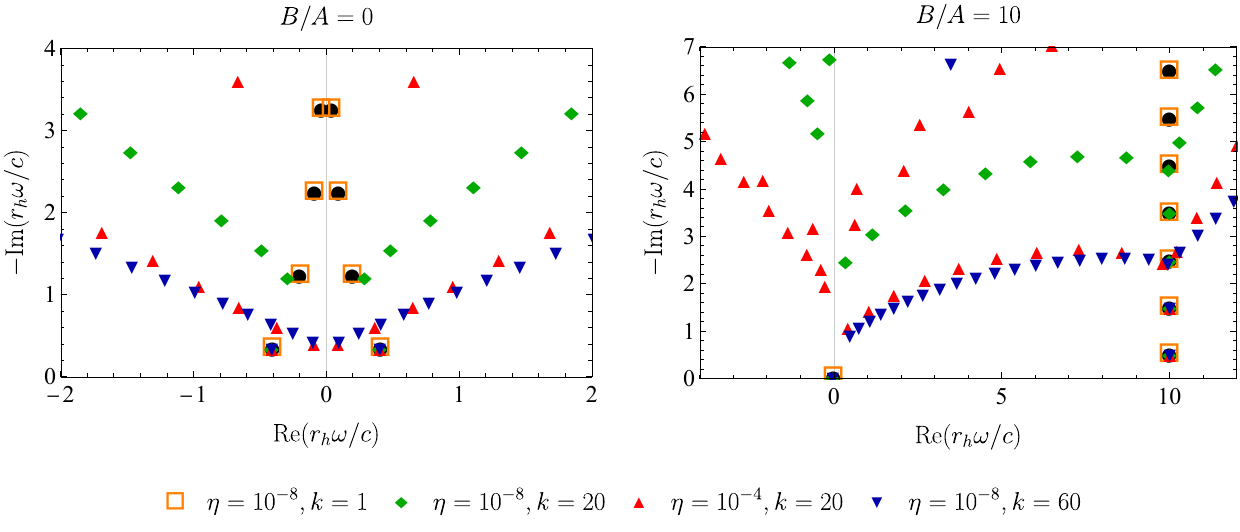}
         \caption{Pertubed QNM spectra for the DBT spacetime. Markers denote the results from different configurations for the sinusoidal perturbations in the potential described by Eq.~\eqref{perturbation}. In both panels, black circles indicate the QNMs of the respective unperturbed system.
         {\em Left panel:} in the stationary case ($B/A = 0$), results are qualitatively the same as in the Schwarzschild spacetime~\cite{rodrigo}. Low amplitudes $\eta$ and wave numbers $k$ do not destabilize the spectrum. As the amplitude or wave number increases, overtones are destabilized, whereas the fundamental mode remains stable. {\em Right oanel:} as rotation increases ($B/A = 10$), the destabilization pattern differs between prograde and retrograde modes. Most importantly, the first prograde overtones exhibit distinct responses: the $n=1$ mode remains entirely unaffected by the perturbations, while the $n=2$ mode is nearly unaltered. This suggests that, in this scenario, these modes share the same stability properties as the fundamental mode.}
		\label{fig4} 	
	\end{figure*}

To explicitly verify the effect of rotation on the spectral instability of the DBT model, we introduce deterministic perturbations into the wave operator \eqref{operatorL}. These perturbations are applied to the effective potential \eqref{effectivepotential} and take the following sinusoidal form,
\begin{equation}
\delta{q}_{rot} = \eta\sin(2\pi k \sigma),
\label{perturbation}
\end{equation}
where $\eta$ represents the perturbation magnitude, which is taken to be small compared to the original potential, and $k$ denotes its wave number.
Despite the \emph{ad hoc} nature of this modification, which does not represent realistic scenarios at a classical level~\cite{Cardoso:2024mrw}, the perturbation \eqref{perturbation} constitutes an effective tool for triggering the spectral instability identified through the pseudospectrum and condition number analysis discussed in the previous section. 

The behavior of QNMs under deterministic perturbations is depicted in Fig.~(\ref{fig4}) for different parameter choices. In particular, we present the results for $(\eta, k) = (10^{-8}, 1)$ in orange squares, the results for $(\eta, k) = (10^{-8}, 20)$ in dark green diamonds, the results for $(\eta, k) = (10^{-4}, 20)$ in upward red triangles, and the results for $(\eta, k) = (10^{-8}, 60)$ in downward dark blue triangles.
The left panel shows the results for a nonrotating analog black hole, indicating qualitative agreement with results from the Schwarzschild spacetime~\cite{rodrigo}. For relatively low perturbation amplitudes $\eta$ and low wave numbers $k$ (orange squares), the spectrum remains stable. However, as the amplitude or the wave number increases, the fundamental mode remains stable while the overtones exhibit the expected instability.
The right panel, on the other hand, considers a high-rotation scenario with $B/A=10$. In this case, the instability pattern changes significantly. Not only do we observe an asymmetry in the instability pattern between prograde and retrograde modes, but we also confirm the predicted trend of overtone stabilization with increasing rotation. Notably, the configurations $(\eta, k) = (10^{-8}, 20)$, $(10^{-4}, 20)$, and $(10^{-8}, 60)$ destabilize all QNMs except for the fundamental mode when $B/A=0$. However, in the high-rotation regime, they no longer affect the first prograde overtone $n=1$, with the $n=2$ prograde overtone remaining nearly unaltered. This behavior can be attributed to the previously observation that, for $B/A=10$, the overtones $n=1$ already occupy the same region of the $\varepsilon$-pseudospectrum as the fundamental mode, whereas $n=2$ is on the verge of entering the same region.
Consequently, they inherit the same stability properties as the fundamental mode, and are not affected by the external perturbation in the potential.

\section{Final remarks}
\label{discussion}
We studied the effect of rotation on the spectral stability of the DBT analog black hole, implementing the hyperboloidal approach to determine the QNMs as solutions of the eigenvalue problem associated with a nonself-adjoint operator. Through the analysis of its pseudospectrum, we were able to identify the potential migration regions of the QNMs, revealing the unstable nature of the system when subjected to external perturbations. The QNMs shift toward the pseudospectrum contour lines, with their redistribution determined by the characteristics of the perturbation. While rotation induces asymmetries in the pseudospectrum, altering the locations of the QNMs in the complex plane, the overall structure of the pseudospectrum remains unaffected. Notably, as rotation increases, the prograde overtones begin to migrate toward the region of the $\varepsilon$-pseudospectrum occupied by the fundamental mode, suggesting that they may share its stability properties in a highly rotating regime. This interpretation is reinforced by the condition number analysis, which shows that the condition numbers of the prograde overtones and of the fundamental mode tend to converge to the same value with increasing rotation.

To test whether rotation enhances spectral stability, we examined the effect of deterministic oscillatory perturbations on the system. In the nonrotating scenario, high frequency perturbations affect the overtones but are not able to change the fundamental mode, as it remains stable across all configurations considered. However, in a highly rotating regime,  specifically $B/A=10$, the first prograde overtone becomes stable under the same ultraviolet perturbations considered in the nonrotating scenario. This behavior suggests that, in fast-rotating systems, prograde overtones may be protected from external perturbations, thus enhancing the reliability of QNMs as indicators of the system's characteristics.

While the DBT model allows for unbounded rotation, the spin of astrophysical black holes, typically modeled by the Kerr metric, possesses a natural upper limit. A recent study on the pseudospectrum of Kerr black holes \cite{Cai:2025irl} analyzed scalar perturbations and confirmed that the instability of the QNM spectrum persists, aligning with our findings. This outcome can be explained by the nonconservative nature of the system remaining significant in the presence of rotation.
Although rotation introduces asymmetries in the pseudospectrum due to QNM delocalization, the extent to which it influences stability in the Kerr case remains an open question, particularly given the constraints on the rotation parameter. Future research could further explore this issue to determine whether rotation can enhance the QNM stability of rapidly rotating Kerr black holes.

Finally, although we focus on \emph{ad hoc} perturbations in this study, the phenomena discussed here can potentially be realized in laboratory settings. Previous works on analog black holes, such as \cite{mauricio3}, have employed vorticity effects to describe physically motivated perturbations to the effective potential that governs wave motion. Recent studies have demonstrated that such perturbations, modeled as Gaussian vortices, could even trigger the instability of the fundamental mode\cite{Malato2025}. Furthermore, a recent study~\cite{Smaniotto:2025hqm}, building on results from~\cite{Solidoro:2024yxi}, has explored how confinement (more generally, boundary conditions) impacts the QNM spectrum of a giant quantum vortex. We note that virtually any property of a gravity simulator that modifies the wave equation can be employed to investigate QNM spectral stability.

\section*{Data Availability}
The data that support the findings of this article are openly available~\cite{data}.

\begin{acknowledgments}
LTP and PHCS would like to thank the Strong
	Group at the Niels Bohr Institute (NBI) for their kind hospitality during the final stages of this work.
	The Tycho supercomputer hosted at the SCIENCE HPC center at the University of Copenhagen was used for supporting this work.
    This study was financed in part by the Coordena\c{c}\~ao de Aperfei\c{c}oamento de Pessoal de N\'{i}vel Superior (CAPES, Brazil) - Finance Code 001, and by the São Paulo Research Foundation (FAPESP), Brasil - Process Numbers 2022/07298-4, 2022/08335-0, 2023/07013-2.   
    MR acknowledges partial support from the Conselho Nacional de Desenvolvimento Científico e Tecnológico (CNPq, Brazil), Grant 315991/2023-2.
	RPM acknowledges support from the Villum Investigator program supported by the VILLUM Foundation (grant no. VIL37766) and the DNRF Chair program (grant no. DNRF162) by the Danish National Research Foundation. 
The Center of Gravity is a Center of Excellence funded by the Danish National Research Foundation under grant No. 184.

\end{acknowledgments}

\bibliography{bh_analog}

\end{document}